\documentclass[useAMS]{mn2e}
\usepackage{graphicx}
\usepackage{lscape}

\title[A new asteroseismological inference of the axion mass]
{The rate of cooling of the pulsating white dwarf star G117$-$B15A:
a new asteroseismological inference of the axion mass}

\author[C\'orsico et al.]
       {A. H. C\'orsico$^{1,2}$, 
        L. G. Althaus$^{1,2}$,
        M. M. Miller Bertolami$^{1,2}$,
        A. D. Romero$^{1,2}$,\newauthor
        E. Garc\'\i a--Berro$^{3,4}$,
        J. Isern$^{4,5}$,
        and             
        S. O. Kepler$^{6}$\\ 
       $^1$Facultad de Ciencias Astron\'omicas
           y Geof\'isicas, Universidad Nacional de La Plata,
           Argentina\\   
       $^2$Consejo Nacional de Investigaciones Cient\'ificas y T\'ecnicas 
          (CONICET), Argentina\\ 
       $^3$Departament de F\'\i sica Aplicada, 
           Universitat Polit\`ecnica de Catalunya,
           c/Esteve Terrades 5, 
           08860 Castelldefels, 
           Spain\\
       $^4$Institute for Space Studies of Catalonia, IEEC,
           c/Gran Capit\`a 2-4, Edif. Nexus 104, 
           08034 Barcelona, 
           Spain\\
       $^5$Institut de Ci\`encies de l'Espai, CSIC, 
           Campus UAB, Facultat de Ci\`encies, Torre C-5, 
           08193 Bellaterra, 
           Spain\\
       $^6$Departamento de Astronomia, Universidade Federal do 
           Rio Grande do Sul, Av. Bento Goncalves 9500
           Porto Alegre 91501-970, RS, Brazil  
}
\begin{document}

\date{}

\maketitle

\label{firstpage}

\begin{abstract}
We employ a state-of-the-art asteroseismological model of G117$-$B15A,
the  archetype of  the H-rich  atmosphere (DA)  white  dwarf pulsators
(also known  as DAV or ZZ  Ceti variables), and use  the most recently
measured value of  the rate of period change for  the dominant mode of
this pulsating star  to derive a new constraint on  the mass of axion,
the  still conjectural non-barionic  particle considered  as candidate
for dark  matter of the  Universe. Assuming that G117$-$B15A  is truly
represented by our asteroseismological  model, and in particular, that
the period of the dominant  mode is associated to a pulsation $g$-mode
trapped in the H envelope, we find strong indications  of the existence 
of extra cooling in this star, compatible with emission of axions of  mass  
$m_{\rm  a}   \cos^2 \beta = \left(17.4^{+2.3}_{-2.7} \right)$~meV.
\end{abstract}


\section{Introduction and context}

Axions are  hypothetical weakly interacting  particles whose existence
was  proposed  about  35  years  ago  as  a  solution  to  the  strong
charge-parity problem in quantum chromodynamics (Peccei \& Quinn 1977;
Weinberg 1978; Wilczek 1978).   They are well-motivated candidates for
dark matter of  the Universe, and their contribution  depends on their
mass (Raffelt 2007),  a quantity that is not given  by the theory that
predicts their  existence.  There are  two types of axion  models: the
KVSZ model  (Kim 1979; Shifman et  al. 1980), where  the axions couple
with  photons and  hadrons, and  the DFSZ  model (Dine  et  al.  1981;
Zhimitskii  1980), where  they  also couple  to  charged leptons  like
electrons.  In  this paper,  we are interested  in DFSZ  axions, those
that interact with electrons.  The coupling strength of DFSZ axions to
electrons  is  defined  through  a  dimensionless  coupling  constant,
$g_{\rm ae}$, which is related to  the mass of the axion, $m_{\rm a}$,
through the relation:

\begin{equation}
g_{\rm ae}= 2.8 \times 10^{-14}\  \frac{m_{\rm a} \cos^2 \beta}{1\ {\rm meV}},
\label{eq1}
\end{equation}

\noindent where  $\cos^2 \beta$  is a free,  model-dependent parameter
that is usually set equal to unity.

Because theory does not place any constraint on the mass of axions, it
must   be  inferred   from   terrestrial  experiments   or  by   using
astrophysical and cosmological arguments.  In particular, stars can be
used to put  constraints on the mass of the  axion (Raffelt 1996).  In
this paper  we focus on white  dwarf stars, which  represent the final
evolutionary  stages  of  low-  and intermediate-mass  stars  ---  see
Althaus  et al.  (2010a)  and references  therein.   White dwarfs  are
excellent  candidates  to test  the  existence  of weakly  interacting
particles, as it  was early recognized by Raffelt  (1986) for the case
of axions.   Because white dwarfs  are strongly degenerate and  do not
have relevant nuclear energy  sources, their evolution is described as
a slow cooling process in which the gravothermal energy release is the
main  energy   source  driving   their  evolution.   At   the  typical
temperatures and  densities found  in the cores  of white  dwarfs, the
emission  of DFSZ  axions is  supposed to  take place  at  the deepest
regions  of  these  stars  through  Compton,  pair  annihilation,  and
bremmsstrahlung  processes, although  the last  mechanism is  the most
relevant one  (Raffelt 1986).  In  this last case, the  axion emission
rate is given by (Nakagawa et al. 1987, 1988):

\begin{equation}
\epsilon_{\rm a}= 1.08 \times 10^{23} \frac{g^2_{\rm ae}}{4\pi} 
\frac{Z^2}{A} T_7^4 F(T, \rho)\ \ [{\rm erg/g/s}]. 
\label{eq2}
\end{equation}

The  mass  of  the  axions  determines how  strongly  they  couple  to
electrons (Eq.~ \ref{eq1}) and then, how large the axion emissivity is
(Eq.~  \ref{eq2}). Since axions  can (almost)  freely escape  from the
interior of  white dwarfs, their existence would  increase the cooling
rate, with  more massive  axions producing larger  additional cooling.
Isern et  al.  (2008, 2009) included axion  emissivity in evolutionary
models of  white dwarfs  and found an  improved agreement  between the
theoretical calculations and  the observational white dwarf luminosity
function.  This  provided the first,  although preliminary, indication
that axions might indeed exist.

Pulsating  DA (namely,  stars with  H-rich atmospheres)  white dwarfs,
also  called ZZ  Ceti or  DAV stars,  are the  most numerous  class of
degenerate pulsators,  with over 148 members  known today (Castanheira
et  al. 2010).   They  are characterized  by multiperiodic  brightness
variations caused  by spheroidal,  non-radial $g$-modes of  low degree
($\ell \leq 2$)  with periods between 70 and 1500  s (Winget \& Kepler
2008; Althaus  et al.  2010a).   The star G117$-$B15A (also  called RY
LMi and  WD 0921$+$354) is  the best studied  member of this  class of
variables.  This star shows oscillation periods $\Pi$ (amplitudes $A$)
of 215.20~s (17.36  mma), 270.46~s (6.14 mma) and  304.05~s (7.48 mma)
that correspond to genuine eigenmodes  (Kepler et al.  1982), and also
shows  the harmonic  of  the  largest amplitude  mode  and two  linear
combinations.   The 215~s mode  is a  $\ell= 1$  mode, as  inferred by
comparing the  UV pulsation amplitude  (measured with the  {\sl Hubble
Space Telescope})  to the optical  amplitude (Robinson et  al.  1995).
The  rate  of  change  of  this  period  with  time  ($\dot{\Pi}\equiv
d\Pi/dt$)  is  very  small,  although  still  detectable,  $\dot{\Pi}
\approx 4 \times  10^{-15}$ s/s (Kepler et al. 2011). In  fact, the stability
of this  period is comparable to  that of the  most stable millisecond
pulsars (Kepler  et al.  2005).   G117$-$B15A has been the  subject of
numerous  asteroseismological analysis, the  more relevant  ones being
those  of Bradley  (1998), C\'orsico  et al.   (2001),  Castanheira \&
Kepler  (2008),  Bischoff-Kim  et  al.   (2008a), and  Romero  et  al.
(2012). In  particular, the approach of Romero  et al.  (2012) allowed
to solve the degeneracy of asteroseismological solutions for G117$-$B15A 
found in previous analysis  by taking into account the predictions of 
full stellar evolution calculations on the structure of DA white dwarfs.

The rates of period change  of pulsation $g$-modes in white dwarfs are
a very important observable  quantity that can yield information about
the core chemical composition.  As a variable white dwarf evolves, its
oscillation periods  vary in response  to evolutionary changes  in the
mechanical structure of the  star. Specifically, as the temperature in
the core of  a white dwarf decreases, the  plasma increases its degree
of  degeneracy  so  the  Brunt-V\"ais\"al\"a frequency  ---  the  most
important physical quantity in  $g$-mode pulsations --- decreases, and
the pulsational spectrum of the star shifts to longer periods.  On the
other hand,  residual gravitational  contraction (if present)  acts in
the opposite  direction, thus  shortening the pulsation  periods.  The
competition  between  the   increasing  degeneracy  and  gravitational
contraction gives  rise to the  detectable temporal rate of  change of
periods.  In particular,  it has been shown (Winget  et al. 1983) that
the rate of change of the  pulsation period is related to the rates of
change  of the  temperature at  the  region of  the period  formation,
$\dot{T}$, and of the stellar radius, $\dot{R}$:

\begin{equation}
\frac{\dot{\Pi}}{\Pi} \approx -a \frac{\dot{T}}{T} + b \frac{\dot{R}}{R}
\label{eq-dotp}
\end{equation}

\noindent where $a$  and $b$ are constants whose  values depend on the
details of the white dwarf  modeling (however, both $a$ and $b \approx
1$).  The first term in Eq.~(\ref{eq-dotp}) corresponds to the rate of
change in period induced by the  cooling of the white dwarf, and it is
a positive  contribution, whereas the second term  represents the rate
of  change due  to gravitational  contraction,  and it  is a  negative
contribution. In  principle, the rate of  change of the  period can be
measured  by  observing a  pulsating  white  dwarf  over a  long  time
interval when one or more very stable pulsation periods are present in
their light curves. In the  case of pulsating DA white dwarfs, cooling
dominates  over gravitational  contraction,  in such  a  way that  the
second  term in  Eq.~(\ref{eq-dotp}) is  usually negligible,  and only
positive values of the observed rate of change of period are expected.

Isern  et al.  (1992)  raised for  the first  time the  possibility of
employing the measured rate of  period change in G117$-$B15A to derive
a constraint  on the  mass of  axions.  By assuming  that the  rate of
change of the  215~s period of G117$-$B15A is  directly related to the
evolutionary timescale  of the star, they considered  the evolution of
DA white dwarfs models with and without axion emissivity, and compared
the  theoretical values of  $\dot{\Pi}$ for  increasing masses  of the
axion with the  observed rate of period change  of G117$-$B15A at that
time.  Employing a semi-analytical treatment they obtained $m_{\rm a}=
8.7$~meV,  assuming  $\cos^2 \beta  =  1$.   Later,  C\'orsico et  al.
(2001) found  $m_{\rm a} \cos^2  \beta \leq 4.4$~meV using  a detailed
asteroseismological  model for  G117$-$B15A.   At the  same time,  the
observational  and theoretical  uncertainties  of the  rate of  period
change were  large enough  that prevented to  be conclusive  about the
necessity of an extra cooling mechanism such as axion emission.  A few
years later,  Bischoff-Kim et al.   (2008b) derived an upper  limit of
$13.5   -   26.5$~meV  for   the   axion   mass   using  an   improved
asteroseismological model for the star and a better treatment of the 
uncertainties involved.

Since the  value of the measured  $\dot{\Pi}$ for the  215~s period of
G117$-$B15A has been changing over the  years and now it seems to have
reached  asymptotically  a  rather   stable  value  around  $4  \times
10^{-15}$ s/s, and since the  modeling of DA white dwarf pulsators has
recently  experienced major  improvements (Althaus  et al.  2010b), we
feel  that  now is  the  right  time to  reanalyze  the  issue of  the
asteroseismological  determination   of  the  axion   mass.   This  is
precisely  the aim  of the present paper. Specifically,  we employ  the very
detailed asteroseismological  model for G117$-$B15A  derived by Romero
et al.   (2012) and  use the  most recent measurement  of the  rate of
change of the  215~s period to set new constraints on  the mass of the
axion.  Preliminary  results of this  research have been  presented in
C\'orsico  et al.   (2011).  The  paper is  organized as  follows.  In
Sect.~2 we give  a succint account of the measurements  of the rate of
period  change in G117$-$B15A,  while in  Sect.~\ref{asteroseismic} we
briefly  present  our  asteroseismological  model of  G117$-$B15A  and
describe in detail some propagation properties of the pulsation modes.
In this section  we also discuss at length  the uncertainties involved
in  our  analysis.   It  follows Sect.~\ref{axion_emission}  where  we
describe the impact that the  inclusion of the axion emissivity in the
asteroseismological    model    has    on   the    pulsation    modes.
Sect.~\ref{axion_mass} is  devoted to place an  improved constraint on
the axion mass.  Finally,  in Sect.~\ref{conclusions} we summarize our
findings and we present our concluding remarks.

\section{Measurements of $\dot{\Pi}$ for the period at 215 s}
\label{observations}

\begin{table}
\centering
\caption{Characteristics  of  G117$-$B15A  as stated by spectroscopy 
and  according  to the  asteroseismological model of Romero et 
al. (2012).}
\begin{tabular}{lcc}
\hline
\hline
 Quantity                        & Spectroscopy      & Asteroseismological            \\
                                 &                   & model                          \\          
\hline
$T_{\rm eff}$ [K]                & $11\,430-12\,500$ & $11\,985 \pm 200$              \\
$M_*/M_{\odot}$                  & $0.530-0.622$     & $0.593 \pm 0.007$              \\
$\log g$                         & $7.72-8.03$       & $8.00\pm 0.09$                 \\
$\log(R_*/R_{\odot})$            &    ---            & $-1.882\pm 0.029$              \\   
$\log(L_*/L_{\odot})$            &    ---            & $-2.497 \pm 0.030$             \\
$M_{\rm He}/M_*$                 &    ---            & $2.39 \times 10^{-2}$          \\
$M_{\rm H}/M_*$                  &    ---            & $(1.25\pm 0.7) \times 10^{-6}$ \\
$X_{\rm C},X_{\rm O}$ (center)   &    ---            & $0.28^{+0.22}_{-0.09} , 0.70^{+0.09}_{-0.22}$ \\
\hline
\end{tabular}\\
{\footnotesize  Note 1: the  ranges of  values in  column 2  have been
  derived  by  taking  into  account  the  spectroscopic  analysis  of
  Robinson et al. (1995), Koester \& Allard (2000), Koester \& Holberg
  (2001), Bergeron et al. (1995, 2004).

  Note 2:  The quoted  uncertainties in the  asteroseismological model
  are the internal errors of our period-fit procedure.  }
\label{table1}
\end{table}

S.  O.  Kepler and collaborators have been observing G117$-$B15A since
1974 to  measure the rate of  period change with time  for the largest
amplitude periodicity  at 215~s.  This is a  challenging endeavor that
requires a huge  investment of telescope time in  order to achieve the
necessary  precision (Winget \&  Kepler 2008).   After about  15~yr of
scrutiny, the first detection of $\dot{\Pi}$ was reported by Kepler et
al.   (1991)  using  observations   performed  with  the  Whole  Earth
Telescope  (WET).   They  found  $\dot{\Pi}=  (12.0  \pm  3.5)  \times
10^{-15}$~s/s,  a  value   substantially  larger  than  the  published
theoretical calculations of  the rate of period change  due to cooling
available at that time,  $\dot{\Pi}= (2-5) \times 10^{-15}$~s/s, which
were  based  on  DA  white  dwarf  models with  pure  C  cores.   This
discrepancy  led to  Isern  et  al.  (1992)  to  postulate that  axion
emission could provide the additional cooling necessary to account for
the large observed rate of period change.

About  a  decade  later,  Kepler  et  al.   (2000)  reported  a  value
$\dot{\Pi}= (2.3 \pm 1.4) \times  10^{-15}$~s/s for the rate of change
of the 215~s period.  This value was about five times smaller than the
value estimated  in 1991, being the  apparent reason a  scatter of the
order  of 1.8~s present  in the  measured times  of maxima  (Kepler et
al. 1995; Costa  et al. 1999).  The value of  $\dot{\Pi}$ of Kepler et
al. (2000) was the value used by C\'orsico et al. (2001) to infer, for
the first time, an asteroseismological upper limit for the axion mass.

In 2005  --- after a total of  31 years of observations  --- Kepler et
al.  (2005)  obtained a  $4 \sigma$ measurement  $\dot{\Pi}_{\rm obs}=
(4.27 \pm 0.80) \times  10^{-15}$~s/s.  Taking into account the proper
motion  effect,  a  rate of  change  of  the  215~s period  with  time
$\dot{\Pi}=  3.57 \pm  0.82 \times  10^{-15}$~s/s was  obtained.  This
value is consistent  with the cooling rate of  white dwarf models only
if cores made of pure C or of a mixture of C and O are considered, and
not with models  in which cores made of heavier  elements are used, as
previously suggested (Kepler et  al. 1990).  This improved measurement
of $\dot{\Pi}$  was used by Bischoff-Kim  et al. (2008b)  to infer new
upper limits for the axion mass.

Very recently,  Kepler et al. (2011) reports  the last measured value  of the
rate  of  period  change,  $\dot{\Pi}_{\rm  obs}=(4.89\pm  0.53)\times
10^{-15}$~s/s.  Applying the proper motion correction, $\dot{\Pi}_{\rm
proper}= (-0.7  \pm 0.2) \times 10^{-15}$  s/s, the rate  of change of
the 215~s period is  $\dot{\Pi}= (4.19 \pm 0.73) \times 10^{-15}$~s/s.
This is the value that we use in the present analysis to constrain the
mass of the axion  --- see Sect.~\ref{axion_mass}.  

Finally, we  would like to point  out at this point  of the discussion
that a  summary of  the different measurements  of $\dot\Pi$,  and the
variations  reported  over the  years  is  graphically illustrated  in
Fig.~1  of  Isern et  al.   (2010).  It  is  quite  apparent that  the
measured  value  of $\dot\Pi$  has  now  stabilized,  being the  jumps
between the  different measurements due  to a number  of observational
artifacts --- see Isern et al. (2010) for further details.

\section{Asteroseismological model for G117$-$B15A}
\label{asteroseismic}

Here,   we  briefly  introduce   the  asteroseismological   model  for
G117$-$B15A, and refer the interested  reader to Romero et al.  (2012)
for  further  details.  Romero et al. (2012)  performed  a detailed  
asteroseismological
analysis of  G117$-$B15A using a  grid of DA white  dwarf evolutionary
models characterized by consistent chemical profiles for both the core
and  the  envelope, and  covering  a  wide  range of  stellar  masses,
thicknesses  of  the  hydrogen  envelope and  effective  temperatures.
These models  were generated with  the {\tt LPCODE}  evolutionary code
--- see,  e.g., Althaus  et al.   (2005) and  references  therein. The
evolutionary calculations  were carried out from the  ZAMS through the
thermally-pulsing and mass-loss phases on  the AGB, and finally to the
domain  of   planetary  nebulae  and  white   dwarfs.   The  effective
temperature, the stellar mass and the mass of the H envelope of our DA
white dwarf models vary within the ranges $14\,000 \ga T_{\rm eff} \ga
9\,000$ K, $0.525  \la M_* \la 0.877 M_{\sun}$,  $-9.4 \la \log(M_{\rm
H}/M_*) \la -3.6$,  where the value of the upper limit of $M_{\rm H}$ is
dependent on $M_*$ and fixed by prior evolution.  
For simplicity, the mass of He  was kept fixed at
the  value  predicted  by   the  evolutionary  computations  for  each
sequence.

\begin{table}
\centering
\caption{The      observed      (G117$-$B15A)     and      theoretical
   (asteroseismological model) periods and rates of period change.}
\begin{tabular}{cccc}
\hline
\hline
$\Pi^{\rm o}$        &  $\Pi^{\rm t}$       & $\ell$ & $k$ \\
\noalign{\smallskip}
$[$s$]$              &   $[$s$]$            &        &     \\  
\hline
  ---                &    189.19            &   1    &  1  \\              
 215.20              &    215.22            &   1    &  2  \\
 270.46              &    273.44            &   1    &  3  \\
 304.05              &    301.85            &   1    &  4  \\
\hline
\hline
$\dot{\Pi}^{\rm o}$  & $\dot{\Pi}^{\rm t}$  & $\ell$ & $k$ \\
\noalign{\smallskip}
$[10^{-15} $s/s$]$   &  $[10^{-15}  $s/s$]$ &        &     \\
\hline
---                  &  3.01                &   1    &  1  \\
$4.19 \pm  0.53$     &  1.25                &   1    &  2  \\  
---                  &  4.43                &   1    &  3  \\  
---                  &  4.31                &   1    &  4  \\  
\hline
\end{tabular}
\label{table2}
\end{table}

In  order to  find an  asteroseismological model  for  G117$-$B15A, 
Romero et al. (2012) sought the model  that minimizes a quality 
function  that measures the
distance between  theoretical ($\Pi^{\rm t}$)  and observed ($\Pi^{\rm
o}$) periods.  The  theoretical periods were assessed by  means of the
pulsation code described  in C\'orsico \& Althaus (2006).   A single  
best-fit  model  with   the  characteristics  shown  in  Table
\ref{table1} was found.   For  the  first  time,  Romero et al. (2012) 
have broken  the  degeneracy  of
solutions for this star  reported in previous studies. This degeneracy
of the solutions involve the  thickness of the H envelope, and depends
on  the   $k$-identification  of   the  three  periods   exhibited  by
G117$-$B15A.  We  found $k=  2$, 3, and  4 for $\Pi^{\rm  o}= 215.20$,
270.46,   and   304.05~s,   respectively,   as   the   only   possible
identification  in the  frame of  our  set of  pulsation models.   The
second  column of  Table~\ref{table1} contains  the ranges  of $T_{\rm
eff}$,   $\log  g$  and   $M_*$  of   G117$-$B15A  according   to  the
spectroscopic studies  of Robinson et  al.  (1995), Koester  \& Allard
(2000), Koester \&  Holberg (2001), Bergeron et al.   (1995, 2004) ---
see  Romero  et  al.    (2012).   The  parameters  characterizing  the
asteroseismological model are shown in column 3.

In Table~\ref{table2} we compare  the observed and theoretical periods
and rates  of change of the  period.  The model  nearly reproduces the
observed periods. This is particularly  true for the case of the 215~s
period.   The  most important  fact  shown  in Table~\ref{table2}  is,
however, that the observed rate of  change of the 215~s period is more
than  3 times  larger than  the theoretically  expected value.   If we
assume  that the  rate  of period  change  of this  mode reflects  the
evolutionary timescale of the  star, then the disagreement between the
observed  and theoretical  values of  $\dot\Pi$ would  be a  hint that
G117$-$B15A  could  be  cooling  faster  than that  predicted  by  the
standard  theory of  white  dwarf evolution.   

\begin{figure} 
\begin{center}
\includegraphics[clip,width=0.9\columnwidth]{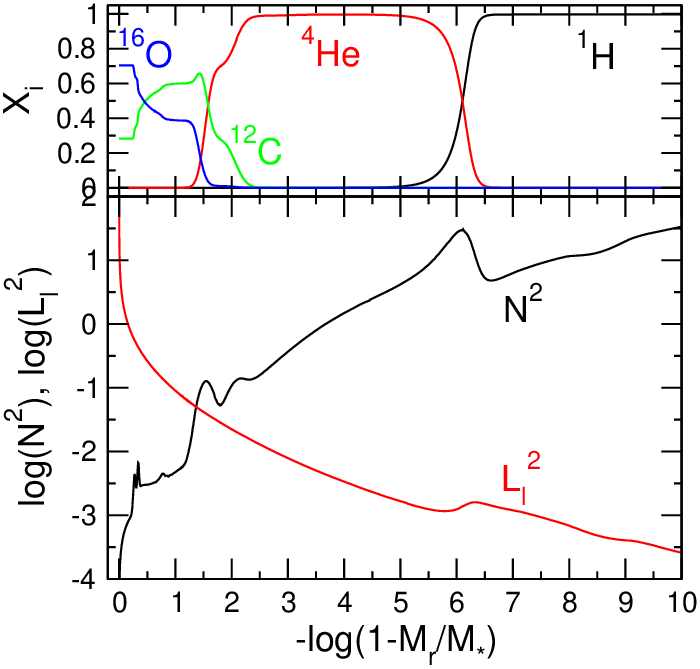} 
\caption{The  internal  chemical stratification  (upper  panel) and  a
  propagation diagram  (lower panel) of  our asteroseismological model
  for G117$-$B15A.}
\label{figure1} 
\end{center}
\end{figure}

The internal  chemical stratification and the  propagation diagram ---
the run  of the  critical frequencies, namely  the Brunt-V\"ais\"al\"a
frequency,  $N$,  and  the  Lamb  frequency,  $L_{\ell}$  ---  of  our
asteroseismological  model  are  shown  in  Fig.~\ref{figure1}.   Each
transition  between   regions  with  different   composition  patterns
produces clear  and distinctive features in $N$,  which are eventually
responsible  for the mode  trapping properties  of the  model (Bradley
1996; C\'orsico et  al. 2002).  In the core  region, there are several
peaks  at  $-\log(q)  \approx   0.4-0.5$  (at  $q  \equiv  1-M_r/M_*$)
resulting  from steep  variations in  the inner  CO profile  which are
caused by  the occurrence  of extra mixing  episodes beyond  the fully
convective core  during central helium burning.  The  extended bump in
$N^2$ at $-\log(q)  \approx 1-2$ is caused by  the chemical transition
of He  to C and O resulting  from nuclear processing in  the prior AGB
and  thermally-pulsing AGB  stages. Finally,  there is  the transition
region between  H and  He at $-\log(q)  \approx 6$, which  is smoothly
shaped by the action of time-dependent element diffusion.

\subsection{Mode trapping}
\label{propa}

\begin{figure} 
\begin{center}
\includegraphics[clip,width=1\columnwidth]{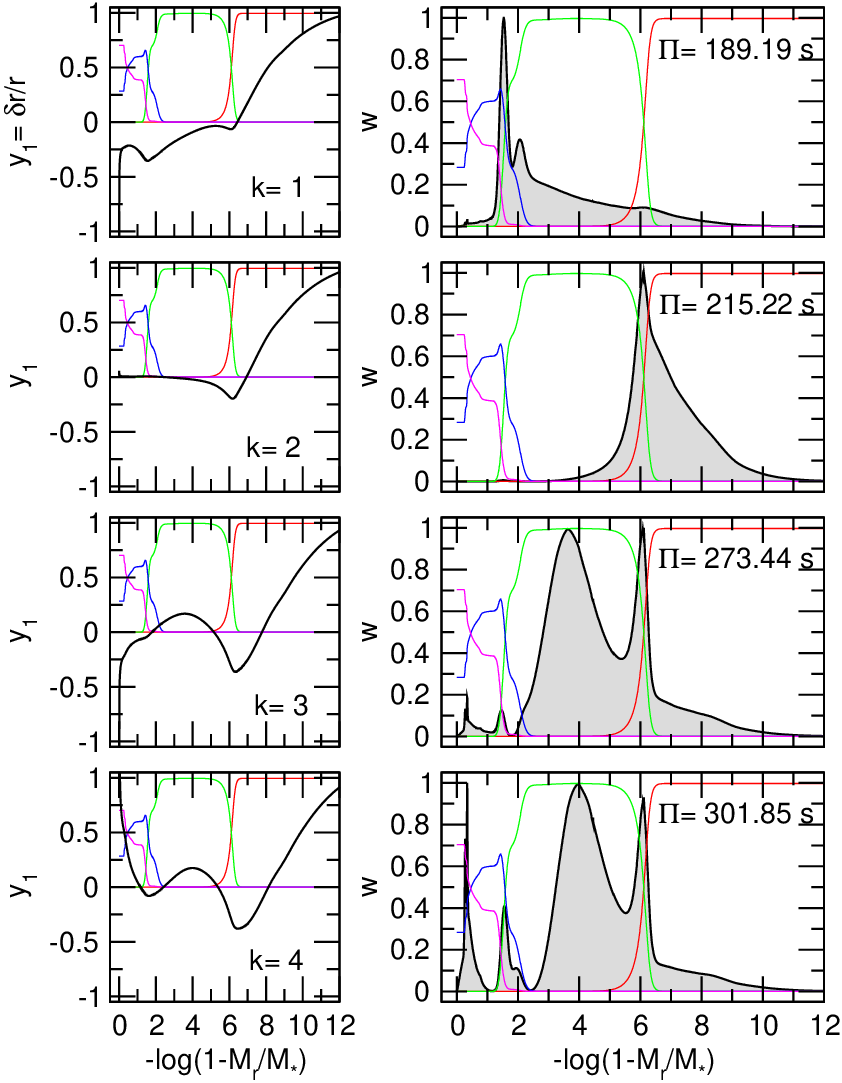} 
\caption{The  radial   eigenfunction  $y_1$  (left   panels)  and  the
  normalized weight  function $w$ (right  panels, shaded) in  terms of
  the outer mass  fraction, corresponding to the modes  with $k= 1, 2,
  3$ and  4 of the  asteroseismological model.  The  internal chemical
  abundances are also included for reference.}
\label{figure2} 
\end{center}
\end{figure}

Table \ref{table2} also reveals that the rate of period change for the
$k= 2$ mode is substantially smaller than for the modes with $k= 1, 3$
and  4.   This  is  because  the modes  have  distinct  mode  trapping
properties.  Briefly, mode trapping  is a mechanical resonance between
the local pulsation wavelength of  a pulsation mode with the thickness
of one  of the compositional layers  (for instance, the  H envelope or
the He buffer).  Mode trapping can reduce the rate of period change of
a mode by up to a factor of 2 if it is trapped in the outer H envelope
(Bradley  1996).  The trapping  properties of  pulsation modes  can be
studied  by examining  their radial  eigenfunction ($\delta  r/r$) and
weight function ($w$) --- see Fig.~\ref{figure2}.  The weight function
of a  given mode  allows to infer  the regions  of the star  that most
contribute to  the period formation (Kawaler et  al.  1985).  Clearly,
$\delta  r/r$  and $w$  for  the mode  with  $k=  2$ have  appreciable
amplitudes only  in the region bounded  by the He/H  interface and the
stellar surface, and  so it is a mode strongly trapped  in the outer H
envelope. This  property for the  $k= 2$ mode  holds also for  all the
models with  structural parameters ($M_*, M_{\rm H},  T_{\rm eff}$) in
the vecinity  of the best-fit  model. Since this mode  is concentrated
closer  to  the  surface,  gravitational contraction  (that  is  still
appreciable in these  regions) acts reducing the period  change due to
evolutionary      cooling      (Bradley     1996). This is  the reason 
why the  rate of
period change of the $k= 2$ mode is small.  In contrast, the remainder
modes are either partially trapped  between the He/H interface and the
He/C/O chemical transition  region (the $k= 1$ mode)  or have non-zero
amplitude through the entire star (the modes with $k= 3$ and 4). Since
these modes  are not affected by gravitational  contraction, they have
larger rates of period change.

The  fact that  the  rate of  period change  for  the $k=  2$ mode  is
somewhat  affected  by gravitational  contraction,  rendering it  less
sensitive to the evolutionary cooling,  could lead us to conclude that
the  mode is  not useful  to  derive constraints  on the  mass of  the
axion. However, it is important to  note that the change of the period
due  to the  increasing  degeneracy resulting  from  cooling is  still
larger than the change due to residual contraction, and so, $\dot{\Pi}
> 0$.  As a  result, the period of the $k= 2$  mode is still sensitive
to cooling and is quite  useful for our analysis.  This statement will
be proven in Sect.~\ref{axion_emission}.

\subsection{Uncertainties in the theoretical value of $\dot{\Pi}$}
\label{errors}

Here, we assess  the uncertainties affecting the value  of the rate of
period  change  for  the   $k=  2$  mode  of  our  asteroseismological
model. This is a crucial  point in order to estimate the uncertainties
in the  derived axion  mass.  We  shall focus on  two main  sources of
errors:  the   poorly  known  $^{12}$C$(\alpha,\gamma)^{16}$O  nuclear
reaction  rate,  and  the  uncertainties  in  the  parameters  of  the
asteroseismological model ($M_*$, $T_{\rm eff}$, and $M_{\rm H}$). 

\subsubsection{The central abundances of carbon and oxygen}
\label{err-reaction}

The $^{12}$C$(\alpha,\gamma)^{16}$O  nuclear reaction plays  a crucial
role in the chemical structure  of the cores of white dwarfs.  Indeed,
the  final CO  stratification of  a  newly born  white dwarf  strongly
depends on the efficiency of  this reaction rate toward the late stage
of core helium burning ---  see, for instance, Althaus et al. (2010a).
In our  evolutionary computations the  rate employed for  the reaction
$^{12}$C$(\alpha,\gamma)^{16}$O is  that of  Angulo et al.  (1999) ---
see Althaus et al. (2010b).   Unfortunately, the rate of this reaction
is not  accurately known.   Table~4  of Kunz et  al. (2002)  gives the
estimated uncertainties of this rate  according to their own study and
the work  by other  authors.  The uncertainties  are between  a factor
$\sim 1.4$ and a factor $\sim 0.6$.

\begin{figure} 
\begin{center}
\includegraphics[clip,width=0.8\columnwidth]{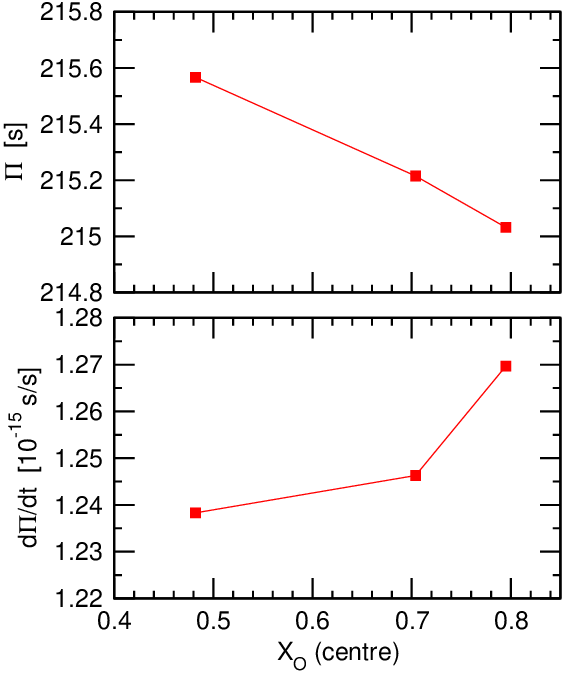} 
\caption{Upper panel: the pulsation period of the mode with $\ell= 1$,
  $k= 2$ in  terms of the central abundance  of $^{16}$O. Lower panel:
  same as upper panel, but for the rate of period change.}
\label{figure3} 
\end{center}
\end{figure}

\begin{table}
\centering
\caption{The central abundances by  mass of $^{12}$C and $^{16}$O for  a
  normal ($f= 1.0$), an enhanced  ($f= 1.5$), and a reduced ($f= 0.5$)
  rate of the $^{12}$C$(\alpha,\gamma)^{16}$O.}
\begin{tabular}{cccc}
\hline
\hline
$f$    & $X_{\rm ^{12}C}$ &  $X_{\rm ^{16}O}$ \\
\hline
 0.5   &  0.505     &   0.482     \\
 1.0   &  0.283     &   0.704     \\
 1.5   &  0.193     &   0.795     \\
\hline
\end{tabular}
\label{table3}
\end{table}

In order  to estimate the  uncertainties of the theroretical  value of
$\dot{\Pi}$ for the $k= 2$ mode introduced by our lack of knowledge of
the  value of  the $^{12}$C$(\alpha,\gamma)^{16}$O  reaction  rate, we
should  compute a  large  number  ($\ga 20\,000$)  of  DA white  dwarf
evolutionary  models like  those computed  in Romero  et  al.  (2012),
employing different  values of this  nuclear reaction rate  to account
for  its uncertainties.   Afterwards, for  each value  of  the adopted
reaction  rate   period  fits  to  G117$-$B15A  to   find  a  suitable
asteroseismological model  should be performed.   Then, the comparison
between  the theoretical  values of  $\dot{\Pi}$ corresponding  to the
various best-fit models should  allow us to estimate the uncertainties
in  the  rate of  period  change induced  by  the  uncertainty in  the
$^{12}$C$(\alpha,\gamma)^{16}$O reaction rate.  At present, due to the
heavy computational load involved,  this procedure is not feasible and
is  far beyond  the  reach of  the  present paper.   Instead, we  have
performed a two-step approximate estimate as described below.

The  first step  consists in  assessing the  variation of  the central
abundances of  C and  O for different  values of the  nuclear reaction
rate.  To this aim,  we computed two additional evolutionary sequences
from the central He burning  stage, assuming an enhanced and a reduced
rate  of the $^{12}$C$(\alpha,\gamma)^{16}$O  reaction.  Specifically,
we multiplied by a  factor $f= 1.5$ and a factor $f=  0.5$ the rate of
Angulo et  al. (1999), thus  comfortably covering the quoted  range of
uncertainty for this rate. In Table~\ref{table3} we show the resulting
central abundance of  C and O in terms of the  adopted rate.  The case
with $f=  1.0$ corresponds to the  standard rate given by  Angulo et al.
(1999).  Note that an increase of  $50 \%$ in the rate translates into
a modest  enhancement of $\sim 12  \%$ in the O  abundance, although a
reduction of $50\%$ in the rate  results in a strong decrease of $\sim
46 \%$ in the  O abundance, being C more abundant than  O in this case
(see Table~\ref{table3}).

In a second step, we estimate  how the theoretical values of $\Pi$ and
$\dot{\Pi}$  of the  asteroseismological model  are affected  when the
central O is  enhanced or reduced.  To do so,  we considered two white
dwarf cooling sequences with different internal CO ratios, but keeping
the  same structure  parameters  as those  of the  asteroseismological
model.  Specifically,  we artificially modified  the O and  C chemical
profiles  of  the asteroseismological  model  at  very high  effective
temperatures ($T_{\rm  eff} \sim 80\,000$ K),  in such a  way that the
unphysical  transitory  associated  to  this procedure  finishes  long
before the  models reach the instability  strip of ZZ  Ceti stars.  The
change  consisted in  replacing  the values  of  $X_{\rm ^{16}O}$  and
$X_{\rm ^{12}C}$ from the stellar centre to the edge of the homogeneus
CO core corresponding  to the case $f= 1$ (normal  rate) by the values
obtained  for $f=  1.5$ (enhanced  rate) and  $f= 0.5$  (reduced rate)
according to Table~\ref{table3}.  After our {\sl ad hoc} procedure, we
allowed  the models  to cool  down  until they  reached the  effective
temperature of  G117$-$B15A.  At this stage, we  computed the periods
and rates of period change.
  
The   results  for   the   mode   with  $k=   2$   are  displayed   in
Fig.~\ref{figure3}.  Due to the fact  that this mode is trapped in the
outer H  envelope, it is almost  insensitive to the details  of the CO
core  chemical  profile  (see  Sect.~\ref{propa}). As  a  result,  its
pulsation period experiences a quite  small decrease of $\sim 0.25 \%$
when the  central $^{16}$O  abundance strongly increases  from $X_{\rm
^{16}O}= 0.482$ to  $X_{\rm ^{16}O}= 0.795$ (upper panel).  This is at
variance with the  periods of the non-trapped modes ($k=  1$, 3 and 4,
not shown in the figure), which  undergo large changes, up to $\sim 21
\%$  in the  case of  the $k  =  4$ mode.   We found  that the  radial
eigenfunction of the $k= 2$ mode remains almost unchanged for the very
different central O  abundances, but the opposite holds  for the modes
with $k=  1, 3$, and 4.   In the lower panel  of Fig.~\ref{figure3} we
depict how the rate of period change  for the $k= 2$ mode changes as a
consequence of  a varying central O abundance.   Not surprisingly, the
value  of $\dot{\Pi}$  for this  mode exhibits  a modest  increase, of
about $2.5  \%$.  Again, this is  because this mode is  trapped in the
outer  envelope.   In  sharp  contrast,  the  other  modes  considered
experience quite large  changes, up to $\sim 60 \%$  for the mode with
$k= 4$.  The increase of the rate of period change for the $k= 2$ mode
for increasing central  O abundances can be readily  explained using a
simple relation  obtained using the Mestel cooling  law (Mestel 1952):
$\dot{\Pi} =  (3-4) \times  10^{-15} (A/14)$ [s/s]  --- see  Kepler et
al. (2005).

The previous  procedure, although not  fully consistent, allows  us to
assess in  an approximate way the  impact of the  uncertainties of the
$^{12}$C$(\alpha,\gamma)^{16}$O reaction rate on the value of the rate
of  change  of  the  mode   with  $k=  2$.  We  adopt  an  uncertainty
$\varepsilon_1 \sim 0.03 \times  10^{-15}$ s/s for $\dot{\Pi}^{\rm t}$
(Fig.~\ref{figure3}).  This is, indeed,  a quite small uncertainty.

\subsubsection{Asteroseismological model}
\label{err-astero}

Another source of error in the theoretical values of $\dot{\Pi}$ comes
from   the  uncertainties   in  the   asteroseismological   model  for
G117$-$B15A.   In this  sense, we  are able  to account  for  the {\sl
internal} errors of the period-fit procedure.  Examining the errors in
the  values of  the stellar  mass, the  effective temperature  and the
thickness of the  H envelope, we estimate that  the uncertainty in the
rate of period change for the  mode with $k= 2$ is $\varepsilon_2 \sim
0.06 \times 10^{-15}$  s/s at most. This value is  twice the error due
to the $^{12}$C$(\alpha,\gamma)^{16}$O reaction rate.


\section{The rates of period change with axion emission}
\label{axion_emission}

In the previous section we have presented results of periods and rates
of period change for G117$-$B15A  that do not take into account other
energy source than gravothermal energy for the evolutionary cooling of
the  star.  We  have found  that  the theoretically  expected rate  of
period change  of the mode  with $k= 2$  is markedly smaller  than the
value  measured  for G117$-$B15A,  suggesting  the  existence of  some
additional cooling mechanism  in this star.  In this  section we shall
assume that this additional cooling  can be entirely attributed to the
emission of axions.

\begin{figure} 
\begin{center}
\includegraphics[clip,width=0.9\columnwidth]{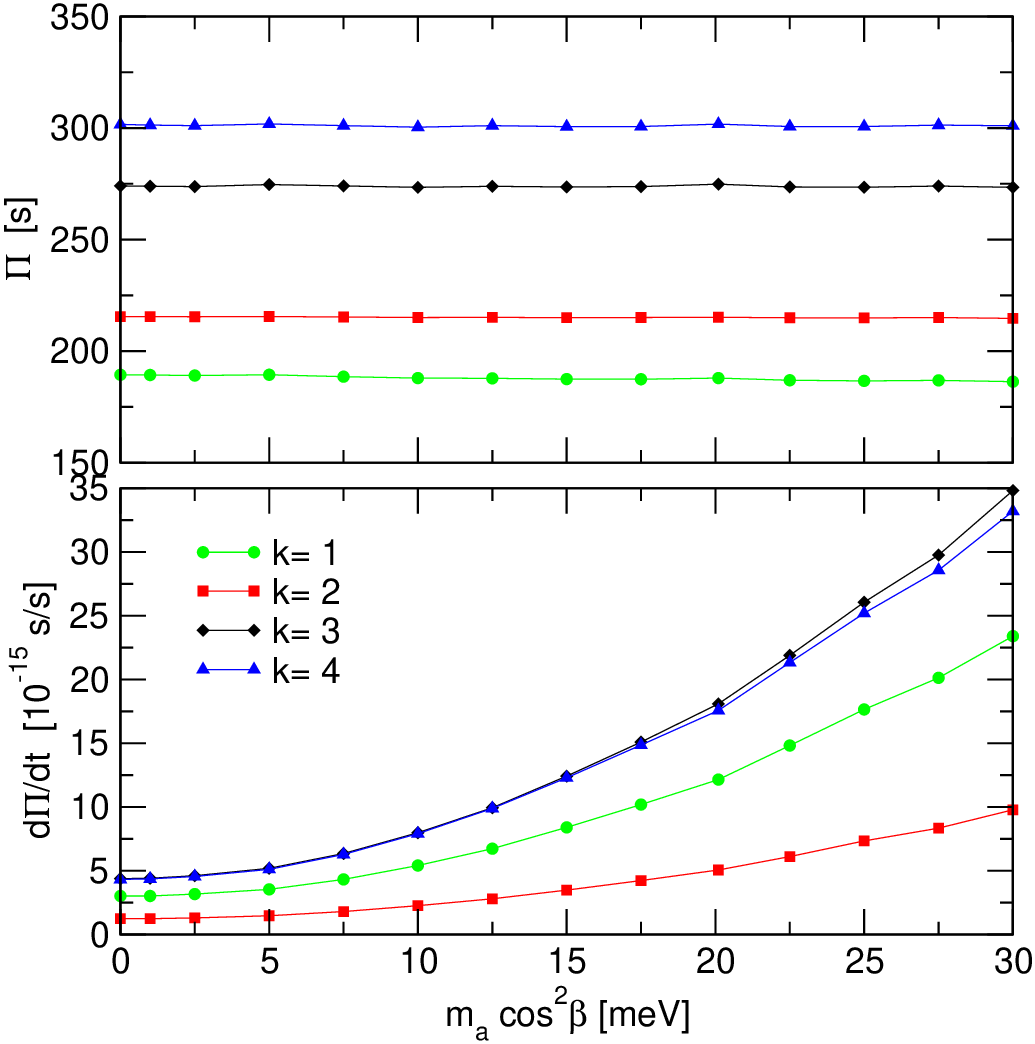} 
\caption{Upper panel:  the pulsation periods of the  modes with $\ell=
  1$, $k=  1, 2,  3$ and 4,  corresponding to  our asteroseismological
  model for G117$-$B15A in terms  of the axion mass. Lower panel: same
  as upper panel, but for the rates of period change.}
\label{figure4} 
\end{center}
\end{figure}

Following C\'orsico et al. (2001), we  have computed a set of DA white
dwarf cooling  sequences incorporating  axion emission. This  has been
done  considering  different  axion  masses and  the  same  structural
parameters ($M_*$, $M_{\rm H}$)  than those of the asteroseismological
model. We have adopted a range of  values for the mass of the axion $0
\leq m_{\rm  a} \cos^2 \beta  \leq 30$ meV,  and we have  employed the
axion  emission rates of  Nakagawa et  al.  (1988).   The evolutionary
calculations  including  the  emission   of  axions  were  started  at
evolutionary stages long  before the ZZ Ceti phase  to ensure that the
cumulative effect of axion  emission have reached an equilibrium value
before  the  models reach  the  effective  temperature of  G117$-$B15A.  
In  our models,  axion  emission occurs  mainly in  the
innermost  regions at $0  \la -\log(q)  \la 1.6$,  reaching a  peak at
$-\log(q)  \sim  0.24$,  although  there  is  also  a  non  negligible
contribution from regions at $1.6\la -\log(q) \la 5$.

The pulsation periods  for the modes with  $\ell = 1$, $k =  1, 2, 3$,
and  4  of the  asteroseismological  model  for  increasing values  of
$m_{\rm a}$ are depicted in the upper panel of Fig.~\ref{figure4}. The
variation of  the periods is negligible,  in spite of  the rather wide
range of axion masses considered.   This result, which was 
first noted by C\'orsico et al.  (2001), implies that due to this additional
cooling  mechanism,  the structure  of  the asteroseismological  model
itself is somewhat affected, but in such a way that, for a
fixed  value of the  effective temperature,  the pulsation  periods are
largely independent of the adopted value of $m_{\rm a}$.

In  the lower  panel of  Fig.~\ref{figure4}  we display  the rates  of
period change for the same  modes. At variance with what happens with
the pulsation periods, the  values of $\dot{\Pi}^{\rm t}$ are visibly affected
by  the  additional   cooling  source,  substantially  increasing  for
increasing values  of $m_{\rm a}$.  In particular, the rate  of period
change  of the mode  with $k=  2$, which  is the  relevant one  in the
present analysis,  increases by a factor  of about 10 in  the range of
axion masses considered. This convincingly demonstrates that, in spite
of the fact  that this mode is less sensitive  to the the evolutionary
cooling of the star due  to its mechanical properties (mode trapping),
it is still  an excellent tool to constrain the mass  of the axion, as
it will be shown below.

\section{Inference of the axion mass}
\label{axion_mass}

\begin{figure} 
\begin{center}
\includegraphics[clip,width=0.9\columnwidth]{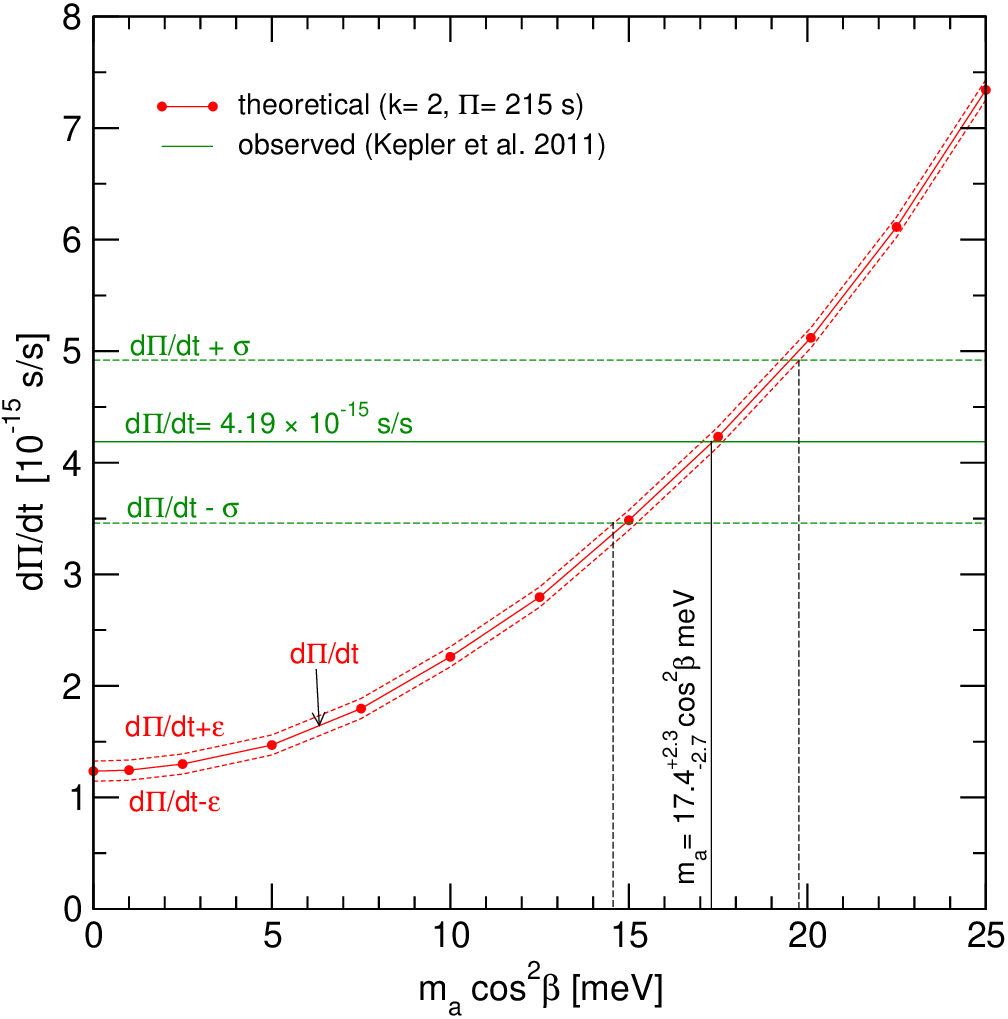} 
\caption{The rate of period change for  the mode with $\ell = 1$, $k =
  2$  of our  asteroseismological model  in  terms of  the axion  mass
  (solid red curve  with dots). Dashed curves represent  the errors in
  $\dot{\Pi}$ due to {\sl  internal} uncertainties in the modeling and
  in  the asteroseismological  procedure. The  horizontal  green lines
  indicate  the observed  value with  its  corresponding uncertainties
  (Kepler et al. 2011).}
\label{figure5} 
\end{center}
\end{figure}

Here,  we  focus  on the  mode  with  $k=  2$,  for  which we  have  a
measurement of  its rate of  period change.  In  Fig.~\ref{figure5} we
display  the theoretical  value  of $\dot{\Pi}$  corresponding to  the
period $\Pi= 215$~s for increasing values of the axion mass (red solid
curve).   The dashed curves  embracing the  solid curve  represent the
uncertainty    in    the    theoretical    value    of    $\dot{\Pi}$,
$\varepsilon_{\dot{\Pi}}=  0.09 \times  10^{-15}$~s/s. This  value has
been obtained  considering the uncertainty  introduced by our  lack of
precise knowledge of the $^{12}$C$(\alpha,\gamma)^{16}$O reaction rate
--- $\varepsilon_1    \sim     0.03    \times    10^{-15}$~s/s,    see
Sect.~\ref{err-reaction}  ---  and  that  due  to the  errors  in  the
asteroseismological   model  ---   $\varepsilon_2  \sim   0.06  \times
10^{-15}$~s/s, see  Sect. \ref{err-astero}.  We are  assuming that the
uncertainty for  the case in  which $m_{\rm a}  > 0$ is the  same than
that computed  for the case in  which $m_{\rm a}= 0$.   If we consider
one standard deviation from  the observational value, we conclude that
the  axion mass  is $m_{\rm  a} \cos^2 \beta= \left(17.4^{+2.3}_{-2.7} 
\right)$ meV.  This value is  about 4 times larger than the upper
limit found by C\'orsico et  al.  (2001), and compatible with the more
recent results of Bischoff-Kim et  al. (2008b). Note that, {\sl if} we
assume  that the  anomalous rate  of  cooling of  G117$-$B15A is  {\sl
entirely} due  to the  emission of axions,  this value is  an indirect
measurement of the mass of the axion, and not just an upper bound.

\section{Discussion and conclusions}
\label{conclusions}

In this paper  we have derived a new  value of the mass of  the (up to
now) elusive  particle called  axion. For this  purpose we used  a new
asteroseismological  model for  G117$-$B15A, an  archetypical  ZZ Ceti
star, derived from fully  evolutionary computations of DA white dwarfs
(Romero et al.   2012), and we employed the  most recent determination
of the  rate of period change  for the largest  amplitude mode ($\ell=
1$, $k= 2$, $\Pi \approx 215$~s) of this star (Kepler et al. 2011).

We first  compared the observed rate  of period change  for this mode,
$\dot{\Pi}=  (4.19 \pm 0.73)  \times 10^{-15}$~s/s,  with the  rate of
period change of our  asteroseismological model, $\dot{\Pi}^{\rm t}= (1.25 \pm
0.09)  \times 10^{-15}$~s/s,  computed under  the assumption  that the
cooling of this  star is governed only by  the release of gravothermal
energy. The fact  that the observed value is more  than 3 times larger
than the  theoretically expected one strongly suggests  that this star
is cooling  faster than the  standard theory of white  dwarf evolution
predicts.  We  have carefully taken into account  the possible sources
of  uncertainties in  the  theoretical  value of  the  rate of  period
change.  We   found  that  the  uncertainties   affecting  the  poorly
determined  $^{12}$C$(\alpha,\gamma)^{16}$O   reaction  rate  have  no
appreciable  impact on  the theoretical  value of  the rate  of period
change, because the mode is largely trapped in the outer H envelope of
the asteroseismological  model. This feature renders  this mode almost
insensitive to  the fine details of  the chemical structure  of the CO
core.  We  also  accounted  for  the  internal  uncertainties  of  the
asteroseismological model, and found that in this case also the errors
are  of modest magnitude,  and substantially  smaller than  the errors
affecting the observed value of $\dot{\Pi}$.

Next  we assumed, following  the idea  first put  forward by  Isern et
al. (1992), that  the additional cooling necessary to  account for the
large observed rate of period  change of G117$-$B15A can be attributed
to axion emission.  Following  C\'orsico et al.  (2001), we introduced
axion emission  in our  asteroseismological model, considering  a wide
range for the axion mass (between $0$ and $30$~meV). We found that the
periods do  not change, but the  rates of period  change are strongly
affected by an increasing value of  the axion mass.  We found that the
mass of  axion necessary  to account for  the observed rate  of period
change  is  $m_{\rm  a}  \cos^2 \beta= \left(17.4^{+2.3}_{-2.7} \right)$~meV, 
where the errors in the axion mass come mainly from errors
in the measurement of the  observed rate of period change.  This value
of the axion mass is substantially larger than the upper limit derived
by C\'orsico et al. (2001), $m_{\rm a} \cos^2 \beta \leq 4.4$~meV, but
still  in  good  agreement  with  the  range  of  values  obtained  by
Bischoff-Kim  et al.   (2008b), $12  \la  m_{\rm a}  \cos^2 \beta  \la
26.5$~meV.   

We must  emphasize at this point  that the conclusion  of our analysis
for  the existence of  an extra  cooling in  G117$-$B15A due  to axion
emission, and the determination of a new value for the axion mass, are
based on  the fact that  our set of  full DA white  dwarf evolutionary
models  predicts that  the  $k= 2$  mode  (215 s  period) is  strongly
trapped  in the  outer H  envelope.  Had  this mode  not  been largely
trapped  in the  outer layers,  higher values  of its  rate  of period
change would  have resulted, thus markedly weakening  the potential of
G117$-$B15A as a tool to constrain  the axion mass.  In this case, the
uncertainties in our derived value for the axion mass should have been
larger. We  can estimate these uncertainties by  considering the value
of the  rate of period change  that the $k=  2$ mode would have  if it
were non-trapped. A  very simple way to do this is  to assume that the
mode should  have a  $\dot{\Pi}^{\rm t}$ similar to  the typical value  of the
rate of period  change of the non-trapped modes with $k=  1, 3$ and 4,
that is, $\langle \dot{\Pi}^{\rm t}_{\rm nt}\rangle \sim 3.9 \times 10^{-15}$
s/s.   Thus, the  value of  $\dot{\Pi}^{\rm t}$  for the  $k= 2$  mode of  the
asteroseismological  model could  be  $\sim 2.7  \times 10^{-15}$  s/s
higher if  it were  a non  trapped mode. By  taking into  account this
uncertainty, we cannot discard the non existence of the axion.

All in  all, we can safety  conclude that, if  the period at 215  s of
G117$-$B15A  is  associated to  a  pulsation  mode  trapped in  the  H
envelope, then the theoretical  models strongly indicate that it would
be necessary  the existence  of an extra  mechanism of energy  loss in
this pulsating white dwarf, consistent with the existence of axions of
$m_{\rm a} \cos^2 \beta \sim 17.4$~meV.  In the context of our full DA
white dwarf  evolutionary models, we found  that the $k= 2$  mode is a
trapped one for all  the models characterized by structural parameters
that  place  them in  the  neighborhoods  of the  asteroseismological
model.  So, the  trapping of this mode in the  H envelope appears very
likely  in the  models,  thus reinforcing  our  conclusion about  the
existence of axions with the quoted mass value.

Besides G117$-$B15A,  the DAV star R548  (ZZ Ceti itself)  is known to
exhibit a  change of  its pulsation period  at $\sim 213$~s  with time
(Mukadam et  al. 2003).  This star  is a potential  objective to study
axions  by   employing  methods  like   the  one  presented   in  this
paper. Based on observations from  1970 to 2007, Mukadam et al. (2009)
have determined a  value of $\dot{\Pi}$ between $(0.8  \pm 1.9) \times
10^{-15}$~s/s and  $(4.3 \pm  1.2) \times 10^{-15}$~s/s,  depending on
the method employed.  Although these values cannot be considered still
as a  {\sl measurement} of the  rate of period change,  it is expected
that a conclusive result in the near future will be obtained.

Isern et al.  (2010) have drawn the attention about the possibility of
employing  also  pulsating DB  (with  He-dominated atmospheres)  white
dwarfs (DBVs) to provide additional constraints to the axion mass.  At
present, there exist two pulsating  DBVs for which is expected to have
available a rate of period change  measured in the next years.  One of
them  is  EC20058$-$5234, a  hot  DBV  star  for which  a  preliminary
estimate of $\dot{\Pi}= 8 \times 10^{-13}$~s/s for the period at 257~s
has  been  reported  (Dalessio  et  al.  2010).   The  other  star  is
KIC~8626021,  the recently  discovered DBV  star in  the  {\sl Kepler}
mission  field   ({\O}stensen  et   al.   2011).   According   to  the
asteroseismological analysis of Bischoff-Kim \& {\O}stensen (2011) and
C\'orsico et al. (2012), this star is  also a hot DBV star. It will be
further monitored with {\sl Kepler}  in the next years, which probably
will allow a measurement of $\dot{\Pi}$.

Clearly,  asteroseismology of pulsating  DA and  DB white  dwarf stars
constitutes a  very exciting avenue  to study particle physics,  and in
particular   axions.  Future   continued  observations   and  possible
measurements of their period drifts will allow to confirm, although in
an indirect way, the existence of axions.


\section*{Acknowledgments}

Part of  this work  was supported by  AGENCIA through the  Programa de
Modernizaci\'on   Tecnol\'ogica    BID   1728/OC-AR,   by    the   PIP
112-200801-00940 grant from CONICET, by MCINN grant AYA2011--23102, by
the ESF EUROCORES Program EuroGENESIS (MICINN grant EUI2009-04170), by
the European  Union FEDER  funds, by the  AGAUR, and  by CAPES/MINCyT.
This research has made use of NASA's Astrophysics Data System.


\label{lastpage}


\begin{thebibliography}{99}

\bibitem{} Althaus, L. G., C\'orsico, A. H., Isern, J., \& Garc\'ia-Berro, E. 2010a, A\&AR, 18, 471 
\bibitem{} Althaus, L. G., C\'orsico, A. H., Bischoff-Kim, A., Romero, A. D., Renedo, I., Garc\'ia-Berro, E., \& Miller Bertolami, M. M. 2010b, ApJ, 717, 897 
\bibitem{} Althaus, L. G., Serenelli, A. M., Panei, J. A. et al. 2005, A\&A, 435, 631 
\bibitem{} Angulo, C., et al. 1999, Nuclear Physics A, 656, 3 
\bibitem{} Bergeron, P., Fontaine, G., Bill{\`e}res, M., Boudreault, S., \& Green, E. M. 2004, ApJ, 600, 404  
\bibitem{} Bergeron, P., Wesemael, F., Lamontagne, R., Fontaine, G., Saffer, R. A., \& Allard, N. F.  1995, ApJ, 449, 258 
\bibitem{} Bischoff-Kim, A., \& {\O}stensen, R. H. 2011, ApJ, 742, L16 
\bibitem{} Bischoff-Kim, A., Montgomery, M. H., Winget, D. E. 2008a, ApJ,  675, 1505 
\bibitem{} Bischoff-Kim, A., Montgomery. M. H., Winget, D. E. 2008b, ApJ,  675, 1512 
\bibitem{} Bradley, P. A. 1998, ApJS, 116, 307 
\bibitem{} Bradley, P. A. 1996, ApJ, 468, 350 
\bibitem{} Bradley, P. A., Winget, D. E., \& Wood, M. A. 1992, ApJ, 391, L33 
\bibitem{} Castanheira, B. G., Kepler, S. O., Kleinman, S. J., Nitta, A., \& Fraga, L. 2010, MNRAS, 405, 2561 
\bibitem{} Castanheira, B. G., Kepler, S. O. 2008, MNRAS, 385, 430 
\bibitem{} Costa, J. E. S., Kepler, S. O., \& Winget, D. E. 1999, ApJ, 522, 973 
\bibitem{} C\'orsico, A. H., Althaus, L. G., Miller Bertolami, M. M., \& Bischoff-Kim, A. 2012, A\&A, 541, 42 
\bibitem{} C\'orsico, A. H., Althaus, L. G., Romero, A. D., Miller Bertolami, M. M., Garc\'ia-Berro, E., \& Isern, J. 2011, ASP proceedings of {\sl ``61st Fujihara seminar: Progress in solar/stellar physics with helio- and asteroseismology''} Ed: Hiromoto Shibahashi, in press (arXiv:1108.354) 
\bibitem{} C\'orsico, A. H., \& Althaus, L. G. 2006, A\&A, 454, 863  
\bibitem{} C\'orsico, A. H., Althaus, L. G., Benvenuto, O. G., \& Serenelli, A. M. 2002, A\&A, 387, 531 
\bibitem{} C\'orsico, A. H., Benvenuto, O. G., Althaus, L. G., Isern, J., Garc\'ia-Berro, E.  2001, New Astr., 6, 197 
\bibitem{} Dalessio, J., Provencal, J. L., Sullivan, D. J., \& Shipman, H. L. 2010, AIP Conference Series, 1273, 536 
\bibitem{} Dine, M., Fischler, W., \& Srednicki, M. 1981, Physics Letters B, 104, 199 
\bibitem{} Isern, J., Garc\'ia-Berro, E., Althaus, L. G., C\'orsico, A. H. 2010, A\&A, 512, A86 
\bibitem{} Isern, J., Catal\'an, S., \& Garc\'ia-Berro, E., 2009, J. of Phys: Conf. Series, 172, 012005 
\bibitem{} Isern, J., Garc\'ia-Berro, E., Torres, S. \& Catal\'an, S., 2008, ApJ, 682, L109 
\bibitem{} Isern, J., Hernanz, M., Garc\'ia-Berro, E. 1992, ApJ, 392, L23 
\bibitem{} Kawaler, S. D., Winget, D. E., \& Hansen, C. J. 1985, ApJ, 295, 547 
\bibitem{} Kepler, S. O. et al. 2011, ASP proceedings of  
{\sl ``61st Fujihara seminar: Progress in solar/stellar physics with 
helio- and asteroseismology''} Ed: Hiromoto Shibahashi, in press 
\bibitem{} Kepler, S. O., Costa, J. E. S., Castanheira, B. G., et al. 2005, ApJ, 634, 1311 
\bibitem{} Kepler, S. O., Mukadam, A., Winget, D. E., et al. 2000, ApJ, 534, L185 
\bibitem{} Kepler, S. O., Winget, D. E., Nather, R.~E., et al. 1995, Baltic Astronomy, 4, 221 
\bibitem{} Kepler, S. O., Winget, D. E., Nather, R. E., et al. 1991, ApJL, 378, L45 
\bibitem{} Kepler, S.~O., Vauclair, G., Dolez, N., et al.\ 1990, ApJ, 357, 204 
\bibitem{} Kepler, S. O., Nather, R. E., McGraw, J. T., Robinson, E. L. 1982, ApJ, 254, 676 
\bibitem{} Kim, J. E.  1979, Physical Review Letters, 43, 103 
\bibitem{} Koester, D., \& Holberg, J. B. 2001, {\sl 12th European Workshop on White Dwarfs}, 226, 299 
\bibitem{} Koester, D., \& Allard, N. F.  2000, Balt Astron 9, 119  
\bibitem{} Kunz, R., Fey, M., Jaeger, M., et al. 2002, ApJ, 567, 643 
\bibitem{} Mestel, L. 1952, MNRAS, 112, 583 
\bibitem{} Mukadam, A. S. et al. 2009, J. of Phys: Conf. Series, 172, 012074
\bibitem{} Mukadam, A. S. et al. 2003, ApJ, 594, 961 
\bibitem{} Nakagawa, M., Adachi, T., Kohyama, Y., \& Itoh, N. 1988, ApJ, 326, 241 
\bibitem{} Nakagawa, M., Kohyama, Y., \& Itoh, N. 1987, ApJ, 322, 291 
\bibitem{} {\O}stensen, R. H., Bloemen, S., Vu{\v c}kovi{\'c}, M., et al. 2011, ApJ, 736, L39 
\bibitem{} Peccei, R.~D., \& Quinn, H. R. 1977, Physical Review Letters, 38, 1440
\bibitem{} Raffelt, G. G.  2007, Journal of Physics A Mathematical General, 40, 6607 
\bibitem{} Raffelt, G. G. 1996, in {\sl Stars as laboratories for fundamental physics : the astrophysics of neutrinos, axions, and other weakly interacting particles} Chicago : University of Chicago Press 
\bibitem{} Raffelt, G. G. 1986, Physics Letters B, 166, 402 
\bibitem{} Robinson, E. L., Mailloux, T. M., Zhang, E., et  al. 1995, ApJ, 438, 908 
\bibitem{} Romero, A. D., C{\'o}rsico, A. H., Althaus, L. G., et al. 2012, 
MNRAS, 420, 1462  
\bibitem{} Shifman, M. A., Vainshtein, A. I., \& Zakharov, V. I.\ 1980, Nucl. Phys. B, 166, 493 
\bibitem{} Weinberg, S. 1978, Physical Review Letters, 40, 223 
\bibitem{} Wilczek, F. 1978, Physical Review Letters, 40, 279 
\bibitem{} Winget, D. E., \& Kepler, S. O. 2008, ARAA, 46, 157 
\bibitem{} Winget, D. E., Hansen, C. J., \& van Horn, H. M. 1983, Nature, 303, 781 
\bibitem{} Zhimitskii, A.P. 1980, Sov. J. Nuc. Phys., 31, 260 

\end{thebibliography}
\end{document}